\documentclass[aps,twocolumn,superscriptaddress,floatfix,letterpaper,nofootinbib]{revtex4-1}

\usepackage{verbatim}

\newcommand{\nn}{\nonumber\\}

\newcommand{\ra}{\rightarrow}

\newcommand{\w}{\mathrm{w}}

\newcommand{\RZ}{{\mathbb{R}/\mathbb{Z}}}
\newcommand\se[1]{\overset{\scriptscriptstyle #1}{=}}

\newcommand\toZ[1]{\lfloor #1 \rfloor}

\begin{document}

\begin{titlepage}

\title{Quantization of Chern-Simons topological invariants\\
for H-type and L-type quantum systems}

\author{Oscar Randal-Williams}

\affiliation{DPMMS,
Centre for Mathematical Sciences,
Wilberforce Road,
Cambridge CB3 0WB,
UK}

\author{Lokman Tsui} 
\affiliation{Department of Physics, Massachusetts Institute of Technology, Cambridge, Massachusetts 02139, USA}

\author{Xiao-Gang Wen} 
\affiliation{Department of Physics, Massachusetts Institute of Technology, Cambridge, Massachusetts 02139, USA}

\begin{abstract} 

In 2+1-dimensions (2+1D), a gapped quantum phase with no symmetry (\emph{i.e.}
a topological order) can have a thermal Hall conductance $\kappa_{xy}=c
\frac{\pi^2 k_B^2}{3h}T$, where the dimensionless $c$ is called chiral central
charge.  If there is a $U_1$ symmetry, a gapped quantum phase can also have a
Hall conductance $\sigma_{xy}=\nu \frac{e^2}{h}$, where the dimensionless $\nu$
is called filling fraction.  In this paper, we derive some quantization
conditions of $c$ and $\nu$, via a cobordism approach to define Chern--Simons
topological invariants which are associated with $c$ and $\nu$.  In particular,
we obtain quantization conditions that depend on the ground state degeneracies on Riemannian surfaces, and quantization conditions that depend on the
type of spacetime manifolds where the topological partition function is
non-zero.

\end{abstract}

\maketitle

\end{titlepage}

\setcounter{tocdepth}{1} 
{\small \tableofcontents }

\section{Introduction}

Different phases of matter are characterized by different orders in
them.\cite{L3726,L3745} Topological order\cite{W8987,WN9077,W9039} is a new
kind of order beyond Landau symmetry breaking order.\cite{L3726,L3745} It
cannot be characterized by the local order parameters associated with the
symmetry breaking, but can be characterized by topological quantum field
theories.\cite{W8951}  Physically, we need to use new topological quantum
numbers to characterize topological orders. In 2+1D, the chiral central charge
$c$\cite{W9125,W9038} (\ie the thermal Hall conductance $\ka_{xy}=c \frac{\pi^2
k_B^2}{3h}T$\cite{KF9732}) of the edge excitations is one such topological
quantum number.  When there is a $U_1$ symmetry, the Hall conductance
$\si_{xy}=\nu \frac{e^2}{h}$ is another such  topological quantum number.  In
other words, the chiral central charge $c$ (partially) characterize 2+1D
topological orders, and the dimensionless Hall conductance $\nu$, together with
$c$,  (partially) characterize 2+1D topological orders with
$U_1$-symmetry.\cite{CGW1038}

In this paper, we like to address the issue of the quantization of those
topological quantum numbers. At first sight, it seems that $c$ and $\nu$ are
not quantized since they do not have to be integers.  On the other hand, $c$
and $\nu$ must be rational numbers, suggesting that they do satisfy certain
quantization conditions which can be complicated.  A quantization condition on
$\nu$ for interacting systems was first obtained by Niu--Thouless--Wu in
\Ref{NTW8572}.  In this paper, we are going to generalize their result and
discuss those complicated quantization conditions that apply to all topological
orders with $U_1$-symmetry.

Before calculating those quantization conditions, we need to distinguish two
types of quantum systems: H-type and L-type.\cite{KW1458} The two
types of quantum systems have different quantization conditions.

The quantum systems in condensed matter physics are all H-type, \ie are all
described by local Hamiltonians defined on smooth spatial manifolds. The
quantum systems in high energy theory and quantum field theory, such as the
topological quantum field theories, are L-type, \ie are described by local
Lagrangian path integrals on smooth spacetime manifolds.  Mathematically, the
L-type topological orders may correspond to unitary fully extended topological
quantum field theories.  The H-type topological orders may correspond to topless
fully extended topological quantum field theories (or $d+\eps$ dimensional
fully extended topological quantum field theories, see
\cite{FETQFT}).

The gapped liquid states\cite{ZW1490,SM1403} with $U_1$ symmetry include
$U_1$-symmetry enriched topological (SET)
orders\cite{CGW1038,LV1334,MR1315,HW1351} and $U_1$-symmetry protected trivial
(SPT) orders.\cite{GW0931,CGL1314,GW1441,LV1219,LW1224,CW1217,SL1204} For 2+1D
H-type gapped states with $U_1$ symmetry, the quantization of $c$ and $\nu$ is
determined from the ground state degeneracy on closed genus $g$ surfaces, and
the main results are given by \eqn{sumB} and \eqn{sumF}.  

In particular, for
H-type invertible gapped states\footnote{By definition, invertible gapped
states have no fractionalized excitations. As a result, they have
non-degenerate ground state on closed smooth spatial manifolds.} with $U_1$
symmetry, $c$ and $\nu$ satisfy the following quantization conditions:
\begin{align}
\label{cnu1}
 \text{bosonic systems: }\
& c ,\ \nu = 0 \text{ mod } 2.
\nonumber\\
 \text{fermionic systems: }\
& c ,\ \nu,\  \frac {\nu-c} 2 = 0 \text{ mod } 1.
\end{align}
The even Hall conductance $\nu=0$ mod 2 for bosonic SPT phases was pointed out
in \Ref{LV1219,LW1224,CW1217,SL1204}.  We remark that the known H-type
invertible topological orders with $U_1$ symmetry do not saturate the above
quantization conditions.  For example, we do not know any  H-type bosonic
invertible topological orders with $c=2$.  Also, we do not know any  H-type
fermionic $U^f_1$ symmetric invertible topological orders with $\nu-c=2$.

For 2+1D L-type gapped states with $U_1$ symmetry, the quantization of $c$ and
$\nu$ is determined by the non-vanishing topological partition functions on
certain types of spacetime manifolds, such as orientable or spin$^\C$.  We find
that, for L-type invertible gapped phases with $U_1$ symmetry, $c$ and $\nu$
satisfy the following quantization conditions:
\begin{align}
\label{cnu2}
 \text{bosonic systems: }\
& 
c = 0 \text{ mod } 8, \ \
\nu = 0 \text{ mod } 2.
\nonumber\\
 \text{fermionic systems: }\
& c ,\ \nu,\  \frac {\nu-c} 8 = 0 \text{ mod } 1.
\end{align}
The above L-type gapped phases with $U_1$ symmetry have a framing anomaly, \ie
the partition functions of the states dependent on the choices of the framing
of the spacetime manifold.\cite{W8951}

If the microscopic Lagrangian path integral is manifestly independent of
choices of the framing (for example, only dependent on the diffeomorphism
equivalent classes of spacetime metrics), then the resulting partition must
also be  independent of choices of the framing.  The resulting phase is said to
be free of framing anomaly. In this case, $c$ must be multiple of
24, and the above quantization conditions reduce to
\begin{align}
\label{cnu3}
 \text{bosonic systems: }\
& c = 0 \text{ mod } 24, \ \ \nu = 0 \text{ mod } 2.
\nonumber\\
 \text{fermionic systems: }\
& c = 0 \text{ mod } 24, \ \ \nu = 0 \text{ mod } 8.
\end{align}

It is interesting to note that the bosonic $E_8$ quantum Hall state with $c=8$
(see \eqn{KmatW} with $K$ given by the $E_8$ matrix \eq{E8}) is an H-type
topological order.  But it cannot be realized by a topological quantum field
theory with no framing anomaly.  In other words,  it cannot be realized by a
Lagrangian path integral that manifestly only depends on the diffeomorphism
equivalent classes of spacetime metrics.  

However, the $E_8$ state can be realized by a Lagrangian path integral with
framing anomaly (\ie depends on the additional framing structure of spacetime
manifold).\cite{W8951}  One such realization is given by the dynamical
Chern--Simons theory \eqn{KmatL} with $K=E_8$.  Here we like to remark that it
is highly nontrivial to define the dynamical Chern--Simons theory \eqn{KmatL} on
spacetime so that we can compute the partition function via a finite
calculation.  For bosonic Chern--Simons theory \eqn{KmatL} (where $K_{II}$ is
even), a non perturbative definition was recently given in \Ref{DW190608270},
by triangulating the spacetime and giving the triangulation a branching
structure. It appears that the branching structure plays a role of framing
structure.

On the other hand, the bosonic $E_8^3$ quantum Hall state (the stacking of
three  bosonic $E_8$ quantum Hall state) can be described by a topological
quantum field theory with no framing anomaly, \ie it can be realized by a
Lagrangian path integral that only depends on the diffeomorphism equivalent
classes of spacetime metrics (\ie does not depend on the framing structure of
spacetime).  An explicit construction of the Lagrangian path integral in terms
of $SO_\infty$ non-linear $\si$-model was given in \Ref{W1477}.

\subsection{Notations and conventions} \label{notation}

We will abbreviate the cup product $a\smile b$ and the wedge product $a\wedge
b$ as $ab$.  We will use $\se{n}$ to mean equal up to a multiple of $n$, and
use $\se{\dd }$ to mean equal up to $\dd f$ (\ie up to a coboundary).  We will
use $\toZ{x}$ to denote the greatest integer less than or equal to $x$, and
$\<l,m\>$ for the greatest common divisor of $l$ and $m$ (with $\<0,m\>\equiv
m$).  

We introduce a symbol $\ft$ to construct fiber bundle $E=F\ft B$ from the fiber $F$
and the base space $B$:
\begin{align}
\text{pt} \to  F \to E \to B\to \text{pt} .
\end{align}
We will also use $\ft$ to construct group extension of $G$ by $N$
\cite{Mor97}:
\begin{align}
1 \to  N \to N\ft_{e_2,\al} G \to G\to 1 .
\end{align}
Here $e_2 \in H^2[G;Z(N)]$ and $Z(N)$ is the center of $N$.  Also $G$ may have
a non-trivial action on $Z(N)$ via $\al: G \to \text{Aut}(N)$.  $e_2$ and $\al$
characterize different group extensions.

Also, we will use $Z_n=\{1,\ee^{\ii \frac{2\pi}{n}},\ee^{\ii
2 \frac{2\pi}{n}},\cdots,\ee^{\ii (n-1) \frac{2\pi}{n}} \}$ to denote an
Abelian group, where the group multiplication is ``$*$''.  We use
$\Z_n=\{0,1,\cdots, n-1\}$ to
denote an integer lifting of $Z_n$, where ``+'' is done without mod-$n$.  In
this sense, $\Z_n$ is not a group under ``+''.  But under a modified equality
$\se{n}$, $\Z_n$ is the $Z_n$ group under ``+''.  Similarly, we will use
$\RZ=[0,1)$ to denote an $\R$-lifting of $U_1$ group.  Under a
modified equality $\se{1}$, $\RZ$ is the $U_1$ group under ``+''.  In this
paper,  we have expressions containing the addition ``+'' of $\Z_n$-valued or
$\RZ$-valued, such as $a^{\Z_n}_1+a^{\Z_n}_2$ where $a^{\Z_n}_1$ and
$a^{\Z_n}_2$ are $\Z_n$-valued.  Those  additions ``+'' are done without mod
$n$ or mod 1.  In this paper, we also have expressions like $\frac1n
a^{\Z_n}_1$.  Such an expression converts a $\Z_n$-valued $a^{\Z_n}_1$ to a
$\RZ$-valued $\frac1n a^{\Z_n}_1$, by viewing the $\Z_n$-value as a $\Z$-value.
(In fact, $\Z_n$ is a $\Z$ lifting of $Z_n$.)

\section{The vector bundle on the moduli space for H-type bosonic systems with
gapped liquid ground states}

In this
paper, we study topological invariants for gapped liquid\cite{ZW1490,SM1403}
ground states without excitations.  In order to characterize different classes
of gapped liquid ground states, we consider a moduli space -- the space of
Hamiltonians with on-site symmetry $G$ that have gapped liquid state as the
ground state.  Here we assume that the Hamiltonians in the moduli space depend
on the metrics $g_{ij}$ of the space as well as the $G$-symmetry twist (\ie the
background $G$-gauge field $A_i$ from gauging the on-site
$G$-symmetry\cite{W1313}). So for a closed $n$-dimensional space $M^n$, with a
certain topology as well as a $G$-principle bundle for the $G$-symmetry twist
on $M^n$, the moduli space is the space of spatial metrics and $G$-connections,
which is denoted as $\cM_{G\ft M^n}$. Here $G\ft M^n$ describes the
$G$-principle bundle:
\begin{align}
 \text{pt} \to G \to G\ft M^n \to M^n \to \text{pt}.
\end{align}
Also \frmbox{we assume that the pairs $(g_{ij},A_i)$ differ by gauge
transformations and diffeomorphisms are equivalent and represent the same point
of the  moduli space $\cM_{G\ft M^n}$.  }

Since on every point of the moduli space, the corresponding Hamiltonian is in a
gapped liquid phase, the subspace formed by the degenerate ground states can be
viewed as the fiber at the point. This way, we obtain a vector bundle over the
moduli space $\cM_{G\ft M^n}$.  It was conjectured in \Ref{W9039} that
\frmbox{the topology of this vector bundle completely characterize the H-type
topological orders, \ie  gapped liquid phases with symmetry.}

In particular, let us consider a closed subspace $B$ in $\cM_{G\ft M^n}$.  The
vector bundle on $\cM_{G\ft M^n}$ reduces to a vector bundle on $B$.  In the
next section, we concentrate on this vector bundle on $B$, and try to relate
the vector bundle to topological term in the partition function of the quantum
system under consideration.  We will follow closely the approach proposed in
\Ref{KW1458,KW200411904}.

\section{Chern--Simons invariants in 2+1D H-type bosonic $U_1$-SET and $U_1$-SPT
orders}

\subsection{Bosonic gapped liquids with $U_1$ symmetry in 2-dimensional space}

In this section, we consider gapped liquids with $U_1$ symmetry in
2-dimensional space for bosonic systems.  Let $\Si_g$ be a $2$-dimensional
closed spatial manifold with a $U_1$ connection, that describes the twisted
global $U_1$ symmetry.  The corresponding $U_1$-bundle is given by $U_1\ft
\Si_g$.  In this case, the moduli space $\cM_{U_1\ft \Si_g}$ of the system is
the space of metrics on $\Si_g$ and the $U_1$-connections.  The degenerate
ground states on $\Si_g$ give rise to a vector bundle over the moduli space
$\cM_{U_1\ft \Si_g}$, where the dimension of the vector is given by the ground
state degeneracy $D_{U_1\ft \Si_g}$.

Let us consider a loop $S^1$ in the  moduli space $\cM_{U_1\ft \Si_g}$.  The
holonomy of the vector bundle around the loop is given by a unitary matrix of
$D_{U_1\ft \Si_g}$ dimension: $W_{S^1}$.  We note that the unitary matrix
$W_{S^1}$ is the non-Abelian geometric phase\cite{WZ8411} of the degenerate
ground states under the adiabatic deformation around the loop $S^1$.

From the  vector bundle, we can obtain a determinant bundle, whose holonomy
around the loop is given by the determinant: Det$(W_{S^1})$.  This phase factor
is directly related to the topological term in the effective action
$S_{eff}=\int_{\Si_g\ft S^1} \dd^3 x\;\cL_{eff}$:
\begin{align}
\label{DetU1}
 \text{Det}(W_{S^1}) = 
\Big( \ee^{\ii \int_{\Si_g\ft S^1}\dd^3 x\; \cL_{eff}} \Big)^{D_{U_1\ft \Si_g} }
\end{align}

The effective action may contain a gravitational Chern--Simons term and a $U_1$
Chern--Simons term:\cite{WZ9253,WZ9200,GF14106812}
\begin{align}
\label{Leff}
& \cL_{eff} \dd^3 x =
-2\pi \frac{c}{24} \om_3 
+ \frac{\nu}{4\pi} (A+s A^{SO_2}) \dd (A+s A^{SO_2})
\nonumber\\
&=
-2\pi \frac{c}{24} \om_3 
+ \frac{\nu}{4\pi} A_{eff} \dd A_{eff} 
\end{align}
where $A$ is the $U_1$ connection 1-form describing the total electromagnetic
field, $s$ is the orbital spin carried by the charged
bosons,\cite{WZ9253,WZ9200} and $A^{SO_2}$ is the time-dependent connection
1-form describing the $SO_2$ tangent bundle of the curved space.
$A_{eff}\equiv A+s A^{SO_2}$ is the effective $U_1$ connection 1-form.  Also
$\om_3$ is the gravitational Chern--Simons term that satisfy $\dd \om_3 = p_1$,
where $p_1$ is the first Pontryagin class of the tangent bundle.  In the above,
$c$ is the chiral central charge of the edge theory, and $\nu$ is proportional
to the Hall conductance:
\begin{align}
\si_{xy}=\frac{\nu}{2\pi}=\nu \frac{e^2}{h}
\end{align}
(in $e=\hbar=1$ unit).

We note that 
$ \frac{\nu}{2\pi} \dd A_{eff} =j$
is the 2-cocycle describing the conserved density and current of charged bosons
in the ground state.  Thus $A_{eff}$ satisfy the quantization condition
\begin{align}
 \int_{\Si_g} \frac{\nu}{2\pi} \dd A_{eff} \in \Z
\end{align}
for the closed 2-dimensional space $\Si_g$.  In fact, $\int_{\Si_g}
\frac{\nu}{2\pi} \dd A_{eff}$ is the number of bosons in the ground state.

Now assume the $S^1$ to be the boundary of a 2-dimension subspace $B^2$ in
$\cM_{U_1\ft \Si_g}$.  \Eqn{DetU1} can be rewritten as\cite{ZWZ8477}
\begin{align}
\label{DetU1Si}
\text{Det}(W_{S^1}) = 
\ee^{D_{U_1\ft \Si_g} \ii 2\pi \int_{\Si_g\ft B^2}
-\frac{c}{24} p_1 + \frac \nu 2 c_1^2 
}, 
\end{align}
where $a =\frac{A_{eff}}{2\pi}$ and $c_1=\dd a$ is the first Chern class.  If
we shrink $S^1$ to a point, $B^2$ becomes a closed 2-dimensional subspace. In
this case, $\text{Det}(W_{S^1})=1$ and
\frmbox{
\begin{align}
\label{DckB}
& D_{U_1\ft \Si_g} \int_{\Si_g\ft B^2}
-\frac{c}{24} p_1 +  \frac \nu 2 c_1^2
 \in \Z ,
\end{align}
for any orientable surface bundle $\Si_g\ft B^2$ and for any $U_1$ bundle
whose first Chern class $c_1$ satisfies $\nu \int_{\Si_g} c_1 \in \Z$.  }
This result allows us to obtain quantization condition of chiral central charge
$c$ and the filling fraction $\nu$.  In particular,
every choice of a surface bundle $\Si_g\ft B^2$ and $U_1$ bundle will give us
a quantization condition of $c$ and $\nu$.  The combination of all those
quantization conditions gives us the strictest constraint on the possible
values of $c$ and $\nu$.  In the following, we will choose some special
combination of surface bundles $\Si_g\ft B^2$ and $U_1$ bundles to obtain
concrete quantization conditions of $c$ and $\nu$.

We note that $\int_{\Si_g\ft  B^2} p_1 =0$ mod 12 for any orientable surface
bundles (also called $\Si_g$-bundles), by \eqref{G4}. If the genus $g$
of the fiber $\Si_g$ is equal or less than 2, then $\int_{\Si_g\ft  B^2}
p_1 =0$.\cite{Meyer} If $g\geq 3$, then we can always find a base
manifold $B^2$ with a genus equal or less than 111, such  that there is a
surface bundle $\Si_g\ft  B^2$ with $\int_{\Si_g\ft  B^2} p_1 =\pm
12$.\cite{E9815} If we choose the $U_1$ bundle to be trivial $c_1=0$,
\eqn{DckB} gives us the quantization condition on $c$ first proposed in
\Ref{KW1458,KW200411904}:
\begin{align}
\label{gsdC}
\frac{c}{2}  D_{U_1\ft \Si_g} \in \Z,  \ \ \
\text{for }  g \geq 3,
\end{align}

Next, we consider a trivial surface bundle with $S^2=\Si_0^2$ as the fiber:
$\Si_g\ft B^2 =\Si_0\times B^2$.  On such a surface bundle, $c_1$ has a
form $c_1=c_1^B+c_1^\Si$ where $c_1^\Si$ lives on $\Si_0$.  We have
\begin{align}
&
\frac12 \nu \int_{\Si_0\times B^2} c_1^2
=\frac12 \nu \int_{\Si_0\times B^2} (c_1^B+c_1^\Si)^2
\nonumber\\
&
=\nu\int_{\Si_0} c_1^\Si \times \int_{B^2} c_1^B 
\end{align}
Therefore, for $U_1$ bundles whose Chern class $c_1$ satisfying $\nu
\int_{\Si_g} c_1 \in \Z$,
$\frac12 \nu \int_{\Si_0\times B^2} c_1^2 $ is always an
integer.  No constraint on $\nu$ is obtained.  Also, $\int_{\Si_0\times B^2}
p_1 =0$.  Therefore, the trivial surface bundle $\Si_g\ft B^2
=\Si_0\times B^2$ does not give us any non-trivial quantization conditions
for $c$ and $\nu$.  So in the following, we will assume $g>0$ for our space
$\Si_g$.

Next, we consider the surface bundle $\Si_g\ft B^2 =S^1_x\times S^1_y\times
S^1_a\times S^1_b$, where $S^1_x\times S^1_y=\Si_1$ is the space and
$S^1_a\times S^1_b=B^2$.  We consider the $U_1$ bundle whose first Chern class
has the following form $c_1 =c_1^{xa} +c_1^{yb}$, where $c_1^{xa}$ lives on
$S^1_x\times S^1_a$ and $c_1^{yb}$ lives on $S^1_y\times S^1_b$.  Since
$\int_{\Si_0\ft B^2} p_1 =0$ and $\int_{\Si_1}c_1^{xa}+c_1^{yb}
=\int_{S^1_x\times S^1_y}c_1^{xa}+c_1^{yb} =0$, the quantization condition
\eqn{DckB} becomes
\begin{align}
&\ \ \ \
\frac{\nu}{2} D_{U_1\ft \Si_1} \int_{S^1_x\times S^1_y\times
S^1_a\times S^1_b} (c_1^{xa}+c_1^{yb})^2 
\nonumber\\
&=
\nu D_{U_1\ft \Si_1} 
\int_{S^1_x\times S^1_a} c_1^{xa}
\int_{S^1_y\times S^1_b} c_1^{yb}
.
\end{align}
We can choose the $U_1$ bundles such that $ \int_{S^1_x\times S^1_a} c_1^{xa}
=0,\pm1$ and $ \int_{S^1_y\times S^1_b} c_1^{yb} =0,\pm1$.  We obtain the
following quantization condition
\begin{align}
 \nu D_{U_1\ft \Si_1} \in \Z .
\end{align}
The above result was first obtained by \Ref{NTW8572}, using a similar
consideration.  Because $\Si_g$ is the connected sum of $g$ $\Si_1$'s, the
above reasoning can be generalized to the case $\Si_g\ft B^2 =\Si_g \times
S^1_a\times S^1_b$, which allows us to obtain
\begin{align}
\label{qck}
 \nu D_{U_1\ft \Si_g} \in \Z ,\ \ \ \ g\geq 1.
\end{align}
This result generalizes that of \Ref{NTW8572}.

To obtain an even stronger result, we consider a surface bundle over surface
$\Si_g\ft \Si_h$.  In Appendix \ref{orw:SurfBun}, we show that, for each
$g\geq 5$, there are many surface bundles $\Si_g\ft \Si_h$ such that each of
them has a 2-cocycle $c_1 \in H^2(\Si_g\ft \Si_h;\Z)$ satisfying
$\int_{\Sigma_g} c_1 =0$. Let $\int_{\Si_g\ft \Si_h} c_1^2 = \eta_c$ and
$\int_{\Si_g\ft  \Si_h} p_1 =3 \eta_p$ where $\eta_c$ and $\eta_p$ are
integers.  We find that the set of allowed $(\eta_p,\eta_c)$ 
is given by (see Appendix  \ref{orw:SurfBun})
\begin{align}
\{(\eta_p,\eta_c) \ | \  \eta_p \in 4\Z, \  \eta_c \in \Z \}.
\end{align}
We obtain a stronger quantization condition on $\nu$ from \eqn{DckB}:
\begin{align}
\label{qck6}
 \frac \nu 2 D_{U_1\ft \Si_g} \in \Z ,\ \ \ \ g\geq 5.
\end{align}
To summarize \frmbox{for a 2+1D bosonic gapped phase with $U_1$ symmetry, its
chiral central charge $c$ and filling fraction $\nu$
satisfy the following quantization conditions
\begin{align}
\label{sumB}
\frac{c}{2}  D_{U_1\ft \Si_g} &= 0 \text{ mod }1 , \ \ \ \ g\geq 3,
\nonumber\\
 \nu D_{U_1\ft \Si_g} &= 0 \text{ mod }1, \ \ \ \ g\geq 1,
\nonumber\\
\frac \nu 2  D_{U_1\ft \Si_g} &= 0 \text{ mod }1 , \ \ \ \ g\geq 5.
\end{align}
where $D_{U_1\ft \Si_g}$ is the ground state degeneracy
on closed space $\Si_g$.
}

\subsection{Examples}

\subsubsection{Bosonic $K$-matrix Abelian topological orders} 

We check $\frac{c}{2}  D_{U_1\ft \Si_g} \in \Z$ for 2+1D bosonic Abelian
topological orders described by the following multi-layer wave function
characterized by symmetric integral matrix $K$ with even diagonal:
\begin{align}
\label{KmatW}
 \Psi(z_i^I)=
\prod_{I,i<j} (z_i^I-z_j^I)^{K_{II}}\hskip -0.5em 
\prod_{I<J,i,j} (z_i^I-z_j^J)^{K_{IJ}}
\ee^{-\sum_{I,i} \frac {|z_i^I|^2}4} ,
\end{align}
where $z_i^I$ the complex coordinates of the $i^\text{th}$ boson in
$I^\text{th}$-layer.  The effective field theory of the above state is
described by $K$-matrix Abelian Chern--Simons theory:
\begin{align}
\label{KmatL}
S_{K} = \int  
\frac{K_{IJ}}{4\pi} a_I \dd a_J + \frac{q_I}{2\pi} A \dd a_I .
\end{align}
where the charge-vector $q_I$ are integers, describing the $U_1$ charge of the
bosons in $I^\text{th}$-layer.  It is believed\cite{WZ9290} that the $K$-matrix
Abelian Chern--Simons theory can realize all the 2+1D Abelian topological
ordors.

After integrating out the matter field $a_I$, we obtain the
effective theory \eqn{Leff} with 
\begin{align*}
c = \text{sgn}(K) = n_+ - n_- ,\ \ \ \ \nu = \v q^\top K^{-1} \v q.
\end{align*}
where $n_{\pm}$ are the number of eigenvalues of $K$ with sign $\pm$. The ground state degeneracy is given by
\begin{align*}
D_{U_1\ft \Si_g}=|\Det K|^g.
\end{align*}

So \eqn{DckB} states that the product $\text{sgn}(K)|\Det K|^g$ is an even integer for $g>2$. To check this, note that $\text{sgn}(K) = n\text{ mod } 2$, where $n$ is the dimension of $K$. So it suffices to check that
\begin{align}
\Det K \in 2\Z ~~~~~~~\text{for odd $n$}.\label{detkeven}
\end{align} To see this, write
\begin{align}
\Det K = \sum_{\si} (-1)^{\si} \prod_{I=1}^{n}K_{I \si(I)} \label{DetK}
\end{align}
which is summed over the permutations $\si$ of $\{1,\dots,n\}$ and $(-1)^{\si}$ is its sign. Note that any $\si$ which involves a diagonal element of $K$ contributes an even term to the sum, so we only need to consider the permutations where $\si(I)\neq I$ for all $I$. Every permutation $\si$ can be expressed as a number of disjoint cycles. The sum of all the lengths of the cycles is $n$, which is odd by assumption. So there exists a cycle whose length is odd. Such cycle cannot have length 1, since we assumed $\si(I)\neq I$. So it has length $\geq 3$. The existence of such a cycle implies $\si \neq \si^{-1}$, since reversing the cycle gives a different permutation. Moreover $\si$ and $\si^{-1}$ contributes equally to the sum \eqref{DetK}, since 
\begin{align*}
\prod_{I=1}^{n}K_{I \si^{-1}(I)} = \prod_{I=1}^{n}K_{\si^{-1}(I) I} = \prod_{I=1}^{n}K_{I\si(I)}.
\end{align*}
So the two terms together is even. So \eqref{DetK} is even.

The conditions on $\nu$, \eqn{qck} and \eqn{qck6}, now become
\begin{align}
&
\Det(K)\; \v q^\top K^{-1} \v q  \in \Z, \ \ 
|\Det(K)|^5\frac{ \v q^\top K^{-1} \v q}{2}  \in \Z, 
\end{align}
for all integral vectors $\v q$. The first expression is satisfied because $\Det(K) K^{-1}$ is an integer matrix. For the second expression, we only need to consider the case when $\Det(K)$ is odd. By \eqref{detkeven} this implies the $n$, the dimension of $K$, is even. So the submatrix of $K$ by deleting the $i$th row and $i$th column has odd dimension for all $i$. Applying \eqref{detkeven} to this submatrix, we see that it has even determinant. Hence the diagonals of the cofactor matrix of $K$ are even, which means $\Det(K) K^{-1}$ has even diagonals. From this, the second expression reads
\begin{align*}
&|\Det(K)|^5\frac{ q_I K^{-1}_{IJ} q_J}{2} \nn
&= \frac{1}{2}|\Det(K)|^5\big(\sum_{I}q^2_I K^{-1}_{II} + \sum_{I\neq J} q_I K^{-1}_{IJ} q_J \big)\nn
&=|\Det(K)|^5\big(\frac{\sum_{I}q^2_I K^{-1}_{II}}{2} + \sum_{I< J} q_I K^{-1}_{IJ} q_J \big)\in\Z.
\end{align*}

From the above result, we find that, for Abelian bosonic topological orders
with $U_1$ symmetry, the filling fraction is quantized as
\begin{align}
 \nu &= \frac{2\times \text{integer}}{\text{gcd}(2D_1, D_1^5 )}
\nonumber\\
&=\begin{cases}
 \frac{1}{D_1}\times \text{integer}, & \text{ if } D_1 = \text{even} \\ 
 \frac{2}{D_1}\times \text{integer}, & \text{ if } D_1 = \text{odd} \\ 
\end{cases}
\end{align}
where $D_1=|\Det(K)|$ is the ground state degeneracy on torus.  In particular,
for bosonic $U_1$-SPT order, $D_1=|\Det(K)|=1$ and $\nu=0$ mod 2.  We note that
the bosonic $\nu=1/m$ Laughlin state has $D_1=m = $ even, and saturates the
above quantization condition.

\subsubsection{Bosonic non-Abelian topological orders described by
$SU(2)_k$ Chern--Simons Theory}

The following wave function
\begin{align}
 \Psi(z_i) = \big(\chi_k(z_i)\big)^2
\end{align}
describes a bosonic non-Abelian topological orders of charge-1 bosons, where
$\chi_k(z_i)$ is the fermion wave function of $k$-filled Landau
levels.\cite{W9102} The low energy effective theory is the $SU(2)_k$
Chern--Simons theory.  The edge is described by a $U_1^{2k}/SU(2)_k$ WZW theory,
whose chiral central charge and filling fraction are\cite{W9102}
\begin{align}
c=2k-\frac{3k}{k+2},\ \ \ \ \nu=\frac{k}{2}.
\end{align}

Using the $SU(2)_k$ Chern--Simons theory, we find that there are $k+1$ anyons in
the bulk labeled by $s=m/2$ where $m=0,1,\dots,k$. The ground state degeneracy
is given by
\begin{align*}
D_{U_1 \ft \Si_g} = \sum_{m} S_{0m}^{-2(g-1)} \in \Z
\end{align*}
where
\begin{align*}
S_{mn} = \sqrt{\frac{2}{k+2}}\sin\big(\frac{(m+1)(n+1)\pi}{k+2}\big)
\end{align*}
So \eqn{sumB} reads
\begin{align}
\label{gsdCsu2}
\frac c2 D_{U_1 \ft \Si_g} = \frac{-3k}{2(k+2)} \big(\frac{k+2}{2}\big)^{g-1} \sum_{m=1}^{k+1}\csc^{2(g-1)}\frac{m\pi}{k+2}\in \Z 
\end{align}
which is verified in Appendix \ref{su2cs} for all integers $k\ge0$ and $g\ge3$.
We also have
\begin{align}
\nu D_{U_1 \ft \Si_g} = \frac k2
 \big(\frac{k+2}{2}\big)^{g-1} \sum_{m=1}^{k+1}\csc^{2(g-1)}\frac{m\pi}{k+2}\in \Z \label{nuD1}
\end{align}
for $g\geq 1$, and
\begin{align}
\frac \nu 2 D_{U_1 \ft \Si_g} = \frac k4
 \big(\frac{k+2}{2}\big)^{g-1} \sum_{m=1}^{k+1}\csc^{2(g-1)}\frac{m\pi}{k+2}\in \Z \label{nuD2}
\end{align}
for $g\geq 5$.
We also verify the above two relations in Appendix \ref{su2cs}.

\subsection{A general point of view}

In general, we may consider a boson system in $d$-dimensional space with an
internal symmetry described by group $G_b$. The bosons may carry non-zero
integer spin and may transform non-trivially under the spatial rotation $SO_d$.
We embed $SO_d$ into $SO_\infty \equiv SO$.  Thus, bosons transform under the
full symmetry group $G_{bSO}=G_b\times SO$.  On curved spacetime with
$G_b$-symmetry twist, we have a $G_{bSO}$ connection $a^{G_{bSO}}$ on the
spacetime.  The connection $a^{G_{bSO}}$ is a special one.  When we project
$G_{bSO}$ to $SO$: $G_{bSO} \xrightarrow{\pi} SO$, the $G_{bSO}$ connection is
projected into an $SO$ connection: $a^{G_{bSO}} \xrightarrow{\pi} a^{SO}$, and
such an $SO$ connection $a^{SO}$ must be the connection for the tangent
bundle of the spacetime.  The low energy effective Lagrangian $L_{eff}$
\eq{Leff} is a Chern--Simons term for group $G_{bSO}$.  This point of view will
help us to understand the effective Lagrangian for fermion system.  Since
$G_{bSO}=G_b\times SO$, the low energy effective Lagrangian $L_{eff}$ \eq{Leff}
is a sum of the gravitational Chern--Simons term and a Chern--Simons term for
group $G_b$.

\section{Topological invariants in 2+1D H-type fermionic enriched topological orders with $U^f_1$ symmetry}

\subsection{Symmetry twist of fermion system}

We have seen that to probe the topological properties of a bosonic gapped
liquid phase with symmetry $G_b$, we can use the symmetry twist described by a
$G_b$-connection $a^{SO}$, plus the curved spacetime described by the
$SO$-connection $a^{SO}$ of the tangent bundle.  In other words, we can use the
$G_{bSO}$-connection $a^{G_{bSO}}$ and effective action $S_{eff}(a^{G_{bSO}})$
to probe the topological properties.  Here, we will discuss the symmetry twist
for fermion systems, which turns out to be mixed with spacetime curvature.

For a fermion system in $d$-dimensional space with an internal symmetry, its
symmetry group $G_f$ has a central $\Z_2^f$ subgroup:
\begin{align}
\label{grpext}
 \text{id}
\to \Z_2^f \to G_f \to G_b \to
 \text{id} .
\end{align}
We also write $G_f = \Z_2^f \ft_{e_2} G_b$, where $e_2$
is a group 2-cocycle 
\begin{align}
 e_2 \in H^2(G_b;\Z_2^f)
\end{align}
describing the group extension \eq{grpext} (see Appendix \ref{cenext}).  The
fermion carry half-integer spins and transform non-trivially under the spatial
rotation $SO_d$, which is enlarged to $SO$.  Thus, fermions transform under the
full symmetry group $G_{fSO}=G_f\ft SO$.  On curved spacetime with
$G_f$-symmetry twist, we have a $G_{fSO}$ connection $a^{G_{fSO}}$ on the
spacetime.  The connection $a^{G_{fSO}}$ is a special one.  When we project
$G_{fSO}$ to $SO$: $G_{fSO} \xrightarrow{\pi} SO$, the $G_{fSO}$ connection is
projected into an $SO$ connection: $a^{G_{fSO}} \xrightarrow{\pi} a^{SO}$, and
such an $SO$ connection $a^{SO}$ must be the connection for the tangent bundle
of the spacetime.  The low energy effective Lagrangian $L_{eff}$ is a
Chern--Simons term for group $G_{fSO}$.  Because $G_{fSO}=G_f\ft SO \neq
G_f\times SO$, the low energy effective Lagrangian $L_{eff}$ \eq{Leff} in
general cannot be written as a sum of the gravitational Chern--Simons term for
group $SO$ and a Chern--Simons term for group $G_f$.
 
There is another way to describe full fermionic symmetry group $G_{fSO}=G_f\ft
SO$: \begin{align} G_{fSO} = (G_f \times Spin)/\Z_2^f \end{align} We note that
the spin group $Spin\equiv Spin_\infty= \Z_2^f\ft_{\w_2} SO$ contains a central subgroup
$\Z_2^f$ such that $Spin/\Z_2^f = SO$, where $\w_2 \in H^2(SO;\Z_2^f)$ is the
group 2-cocycle describing the group extension \begin{align} \text{id} \to
\Z_2^f \to Spin \to SO \to \text{id} .  \end{align} $G_f=\Z_2^f\ft_{e_2} G_b$
also contains a central subgroup $\Z_2^f$ such that $G_f/\Z_2^f =G_b$.
$G_{fSO}$ is obtained by identify these two $\Z_2^f$ subgroups in $G_f \times
Spin$.  Let us described the $G_{fSO}$-connection using this point of view.

We first triangulate the spacetime, where the vertices are labeled by
$i,j,k,\cdots$, the links labelled by $ij,jk,\cdots$, \etc.  A $G_f$-connection
on the triangulation is described by the group elements of $G_f$ on the links:
$a^{G_f}_{ij} \in G_f$.  We can view $G_f=\Z_2^f \ft_{e_2} G_b$ and use the
pair $(a^{G_b}_{ij}, a^{\Z^f_2}_{ij})$ to label the group elements of $G_f$,
where $a^{G_b}_{ij} \in G_b$ and $a^{\Z^f_2}_{ij} \in \Z_2^f$.  Similarly, a
$Spin$-connection on the triangulation is described by the group elements of
$Spin$ on the links: $a^{Spin}_{ij} \in Spin$.  We can view $Spin=\Z_2^f
\ft_{\w_2} SO$ and use the pair $(a^{SO}_{ij}, \t a^{\Z^f_2}_{ij})$ to label
the group elements of $Spin$, where $a^{SO}_{ij} \in SO$ and $\t
a^{\Z^f_2}_{ij} \in \Z_2^f$.  Now, the $G_{fSO}$-connection is obtained by
requiring $\t a^{\Z^f_2}_{ij}= a^{\Z^f_2}_{ij}$.  Also the $SO$-connection
$a^{SO}_{ij}$ describes the tangent bundle of the spacetime.

Now, let us describe a nearly flat $G_{fSO}$-connection that describe the
curved spacetime and twisted symmetry.  The nearly flat condition is given by
\begin{align}
 a^{G_{fSO}}_{ij} a^{G_{fSO}}_{ij}  \approx a^{G_{fSO}}_{ik}
\end{align}
for all the triangles $ijk$.  It implies that the $G_f$-connection is nearly
flat
\begin{align}
( a^{G_b}_{ij} , a^{\Z^f_2}_{ij}) \circ ( a^{G_b}_{jk} , a^{\Z^f_2}_{jk})
\approx ( a^{G_b}_{ik} , a^{\Z^f_2}_{ik}),
\end{align}
and the $Spin$-connection is nearly flat
\begin{align}
( a^{SO}_{ij} , a^{\Z^f_2}_{ij}) \circ ( a^{SO}_{jk} , a^{\Z^f_2}_{jk})
\approx ( a^{SO}_{ik} , a^{\Z^f_2}_{ik}).
\end{align}
In other words (see Appendix \ref{cenext})
\begin{align}
 a^{G_b}_{ij}  a^{G_b}_{jk} &\approx  a^{G_b}_{ik} ,
\nonumber\\
 a^{SO}_{ij}  a^{SO}_{jk} &\approx  a^{SO}_{ik} ,
\end{align}
and
\begin{align}
 a^{\Z^f_2}_{ij} + a^{\Z^f_2}_{jk} 
-a^{\Z_2^f}_{ik} &\se{2}  e_2(a^{G_b}_{ij},a^{G_b}_{jk}) =(e_2)_{ijk},
\nonumber\\
 a^{\Z^f_2}_{ij} + a^{\Z^f_2}_{jk} 
-a^{\Z_2^f}_{ik} &\se{2}  \w_2(a^{SO}_{ij},a^{SO}_{jk}) =(\w_2)_{ijk} .
\end{align}
The above can be rewritten as
\begin{align}
 \dd  a^{\Z^f_2} \se{2} e_2(a^{G_b}), \ \ \ \
 \dd  a^{\Z^f_2} \se{2} \w_2(a^{SO}).
\end{align}
Since $a^{SO}$ is the connection of the tangent bundle of the spacetime,
$\w_2(a^{SO})=\w_2$ is the second Stiefel--Whitney class of the  tangent bundle
of the spacetime.  We see that the $G_b$-connection and the $SO$-connection on
spacetime are not arbitrary.  They must satisfy the constraint
\begin{align}
e_2(a^{G_b}_{ij},a^{G_b}_{jk}) \se{2}  \w_2(a^{SO}_{ij},a^{SO}_{jk}) =\w_2.
\end{align}
In other words, the nearly flat $G_b$ bundle on the spacetime (describing the
symmetry twist) is not arbitrary.  Its nearly flat connection $a^{G_b}$ must
satisfy
\begin{align}
e_2( a^{G_b} ) \se{2,\dd} \w_2.
\end{align}

\subsection{$U^f_1$ symmetry and spin$^\C$ structure}

Let us consider an example of the above result.  A fermion system with fermion
number conservation has a $U^f_1$ symmetry.  Assume all the odd charges of
$U^f_1$ are fermionic, then $U^f_1$ has a $\Z_2^f$ subgroup generated by the
$\pi$-rotation, which corresponds to the fermion-number-parity.  
It is more convenient to view $U^f_1$ as a group extension of $U_1$ by
$\Z_2^f$, \ie $U^f_1=\Z_2^f\ft_{e_2} U_1$ where $e_2$ is the $\Z_2$ valued
group 2-cocycle that generates $ H^2(U_1,\Z_2)$ (see Appendix \ref{cenext}).  

Using $\RZ$-valued $a^{(\RZ)^f}$ to label group elements in $U^f_1$,
$\RZ$-valued $a^{\RZ}$ to label group elements in $U_1$, and $\Z_2$-valued
$a^{\Z_2^f}$ to label group elements in $\Z_2^f$, we find that
\begin{align}
\label{U1fU1}
  a^{(\RZ)^f} =
\frac12  a^{\RZ}+\frac12  a^{\Z_2^f},
\end{align}
Here $\RZ=[0,1)$  and $\Z_n=\{0,1,\cdots,n-1\}$.
Now the group multiplication in $U^f_1$ can be rewritten as
\begin{align}
  a^{(\RZ)^f}_1 & + a^{(\RZ)^f}_2
 =
\frac12  a^{\RZ}_1+\frac12  a^{\RZ}_2
+\frac12  a^{\Z_2^f}_1 +\frac12  a^{\Z_2^f}_2
\nonumber\\
&= \frac12 ( a^{\RZ}_1+ a^{\RZ}_2-\toZ{ a^{\RZ}_1+ a^{\RZ}_2})
\nonumber\\
&\ \ \ \ \
+(\frac12  a^{\Z_2^f}_1+\frac12  a^{\Z_2^f}_2+\toZ{ a^{\RZ}_1+ a^{\RZ}_2})
\nonumber\\
&= \frac12 ( a^{\RZ}_1+ a^{\RZ}_2-\toZ{ a^{\RZ}_1+ a^{\RZ}_2})
\nonumber\\
&\ \ \ \ \
+(\frac12  a^{\Z_2^f}_1+\frac12  a^{\Z_2^f}_2
+  e_2( a^{\RZ}_1, a^{\RZ}_2) ),
\end{align}
where $\toZ{x}$ denotes the largest integer smaller than or equal to $x$,
and
\begin{align}
  e_2( a^{\RZ}_1, a^{\RZ}_2) =\toZ{ a^{\RZ}_1+ a^{\RZ}_2}
\end{align}
which is a $\Z_2$-valued group 2-cocycle $e_2\in H^2(U_1;\Z_2)$ that
characterizes the group extension of $U_1$ by $\Z_2^f$:
\begin{align}
U_1^f = \Z_2^f\ft_{e_2} U_1. 
\end{align}
Note that $e_2( a^{\RZ}_1, a^{\RZ}_2)$ is a smooth function of $ a^{\RZ}_1,
a^{\RZ}_2$ when $ 0< a^{\RZ}_1 <\frac12$ and $0< a^{\RZ}_2<\frac12$.  But it
has discontinuities in other places.  It turns out that 
\begin{align} e_2
\se{2,\dd} c_1^{U_1} 
\end{align} where $c_1^{U_1}\in H^2(U_1,\Z) $ is the first
Chern class of $U_1$ viewed as group 2-cocycle.

The ``Chern Class'' for $U^f_1$ and the Chern Class for $U_1$ are related (see
\eqn{U1fU1})
\begin{align}
 c_1^{U_1^f} = \frac12  c_1^{U_1} , 
\end{align}
where $c_1^{U_1} = \dd a^{\RZ}$ and $c_1^{U_1^f} = \dd a^{(\RZ)^f}$.  We like
to point out that the $U^f_1$-connection $a^{(\RZ)^f}$ corresponds to the
electromagnetic gauge potential plus the connection of the curved space:
\begin{align}
  A_{eff} = 2\pi a^{(\RZ)^f} ,
\end{align}
and the Chern class $c_1^{U_1^f}$ corresponds to
the field strength of the effective $U^f_1$-connection
\begin{align}
 F_{eff} =\dd A_{eff} = 2\pi c_1^{U_1^f} .
\end{align}
We find that for fermion system with $U^f_1$ symmetry, the
Chern number on a closed 2-dimension space $C^2$ satisfies
\begin{align}
 \int_{C^2} c_1^{U_1^f} =\frac1{2\pi} \int_{C^2}  F_{eff} \se{1}
\frac12 \int_{C^2} \w_2 .
\end{align}
In other words, on a spin manifold $\w_2\se{2,\dd} 0$, the Chern number
$\frac1{2\pi} \int_{C^2}  F_{eff}$ is quantized as integers.  On non-spin
manifold, the fermion system with $U^f_1$ symmetry can still be defined,
provided that the Chern number $\frac1{2\pi} \int_{C^2}  F_{eff}$ is quantized
as half-integers when $\int_{C^2} \w_2 \se{2} 1$.

We see that for fermion systems, there is a constraint on the $U^f_1$
connection $a^{U^f_1}=\frac1{2\pi}A_{eff}$ and the curved spacetime.  When $
A_{eff}=0$, this  constraint requires that $\w_2\se{2,\dd}0$, \ie the spacetime
to be spin.  This leads to the general impression that the appearance of
fermions requires the spacetime manifold to be spin.  Here we see that  the
appearance of $U^f_1$ fermions requires the spacetime manifold to be spin only
when there is no background effective $U^f_1$ gauge field.  In the presence of
effective background $U^f_1$ gauge field, the spacetime may not be spin.  This
result can be summarized more precisely by the following statement: the
appearance of $U^f_1$ fermions requires the spacetime manifold to be spin$^\C$,
and the $U^f_1$ connection $a^{(\RZ)^f}$ is a spin$^\C$ structure.

We know that an orientable manifold is spin if and only if its
$\w_2\se{2,\dd}0$.  Similarly, an orientable manifold is spin$^\C$ if and only
if $ \bt_2 \w_2 \se{\dd}0$.  Here $\bt_2 = \frac12 \dd $ is the the Bockstein
homomorphism.  In order words, an orientable manifold $M$ is spin$^\C$ if and
only if its $\w_2$ is the mod 2 reduction of a $\Z$-valued 2-cocycle in
$H^2(M;\Z)$.  Also a complex vector bundle of rank $n$ can be viewed as a real
vector bundle of rank $2n$. The $n^\text{th}$ Chern class $c_n^U$ of the
complex vector bundle and the  $2n^\text{th}$ Stiefel--Whitney class are
related
\begin{align}
 \w_{2n} \se{2,\dd} c_n^U.
\end{align}
This implies that a complex manifold is always spin$^\C$.

\subsection{Fermionic gapped liquids with $U^f_1$ symmetry in 2-dimensional
space}

Now, we consider a gapped liquids with $U^f_1$ symmetry in 2-dimensional space
for fermionic systems.  The effective action may contain a gravitational
Chern--Simons term and a $U^f_1$ Chern--Simons term:
\begin{align}
\label{LeffF}
\cL_{eff} \dd^3 x =
-2\pi \frac{c}{24} \om_3 
+ \frac{\nu}{4\pi} A_{eff}\dd A_{eff}
\end{align}
where $A_{eff}$ is the $U^f_1$ connection 1-form describing the total
electromagnetic field, plus the connection from the curved space due to the
orbital spin of the fermions.  Follow the same discussion for the bosonic case,
we find that the quantization conditions of $c$ and $\nu$ for fermionic case can
be obtained from (see \eqn{DckB})
\frmbox{
\begin{align}
\label{DckF}
& D_{U_1\ft \Si_g} \int_{\Si_g\ft B^2}
-\frac{c}{24} p_1 + \frac \nu 8  (c_1^{U_1})^2
 \in \Z ,
\end{align}
for any orientable surface bundle $\Si_g\ft B^2$ and for any $U_1$ bundle whose
first Chern class $c_1^{U_1}$ satisfies $\frac12 \nu \int_{\Si_g} c_1^{U_1} \in \Z$ and
$c_1^{U_1} \se{2,\dd} \w_2$.  } Here $c_1^{U_1} = \dd a^{\RZ}$,
$a^{\RZ}=2\frac{A_{eff}}{2\pi}$ is the $U_1$-connection for charge-$2$ bosons
(fermion pairs), and $\w_2$ is the second Stiefel--Whitney class of the tangent
bundle of $\Si_g\ft B^2$.

Next, we choose some special surface bundles $\Si_g\ft B^2$ and the allowed
$U_1$ bundles to obtain quantization conditions of $\nu$ and $c$.  Let us
choose the surface bundle to be $\Si_g\ft B^2 =S^1_x\times S^1_y\times
S^1\times S^1$, where $S^1_x\times S^1_y$ is the space and $S^1\times S^1=B^2$.
The second Stiefel--Whitney class of such a surface bundle is trivial $\w_2
\se{2,\dd} 0$.  Thus the allowed $U_1$ bundles satisfy $\int_{C^2} c_1^{U_1} =$
even integers.  We consider the following form pf $c_1^{U_1} =c_1^x +c_1^y$,
where $c_1^x$ lives on $S^1_x\times S^1$ and $c_1^y$ lives on $S^1_y\times
S^1$.  Since $\int_{\Si_0\ft B^2} p_1 =0$ and $\int_{S^1_x\times
S^1_y}c_1^x+c_1^y =0$, the quantization condition \eqn{DckF} becomes
\begin{align}
&\ \ \ \
\frac \nu 8 D_{U_1\ft \Si_1} \int_{S^1_x\times S^1_y\times
S^1\times S^1} (c_1^x+c_1^y)^2 
\nonumber\\
&=
\frac \nu 4 D_{U_1\ft \Si_1} 
\int_{S^1_x\times S^1} c_1^x
\int_{S^1_y\times S^1} c_1^y
.
\end{align}
We can choose the $U_1$ bundle such that $ \int_{S^1_x\times S^1} c_1^x
=0,\pm2$ and $ \int_{S^1_y\times S^1} c_1^y =0,\pm2$.
We obtain the following quantization condition
\begin{align}
 \nu D_{U_1\ft \Si_1} \in \Z .
\end{align}
The above result was first obtained by \Ref{NTW8572}.  Again, because $\Si_g$,
$g\geq 1$,
is the connected sum of $g$ $\Si_1$'s, the above reasoning can be generalized
to the case $\Si_g\ft B^2 =\Si_g \times S^1\times S^1$ whose second
Stiefel--Whitney class is still trivial $\w_2 \se{2,\dd} 0$, which allow us to
obtain
\begin{align}
 \nu D_{U_1\ft \Si_g} \in \Z ,\ \ \ \ g\geq 1.
\end{align}

If $\Si_g\ft B^2$ is spin, then $\int_{\Si_g\ft B^2} p_1 =0$ mod 48 by
\eqref{G4bis}. We explain in Appendix \ref{orw:SurfBun} that as long as $g \geq
9$ there is such a spin surface bundle with  $\int_{\Si_g\ft B^2} p_1 =\pm 48$,
in which case we may choose the $U_1$-bundle to be trivial.  This gives us a
quantization of $c$ (see \Ref{KW200411904}):
\begin{align}
\label{gsdCF}
2 c  D_{U_1\ft \Si_g} \in \Z,  \ \ \ \text{for }  g \geq 9.
\end{align}

To obtain additional quantization condition, we consider a more general surface
bundle over surface $\Si_g\ft B^2$.  In Appendix \ref{orw:SurfBun}, we show
that, for each $g\geq 5$, there are many surface bundles $\Si_g\ft B^2$
equipped with a 2-cocycle
$c_1^{U_1} \in H^2(\Si_g\ft B^2;\Z)$ satisfying (1) $\int_{\Sigma_g} c_1^{U_1}
=0$ and (2) $c_1^{U_1}\se{2,\dd} \w_2$.  Let $\int_{\Si_g\ft B^2} (c_1^{U_1})^2
= \eta_c$ and $\int_{\Si_g\ft B^2} p_1 =3 \eta_p$ where $\eta_c$ and $\eta_p$
are integers.  \Eqn{DckF} becomes
\begin{align}
\label{etaZ}
D_{U_1\ft \Si_g} (-\eta_p \frac{c}{8}  + \eta_c\frac \nu 8 ) \in \Z, 
\end{align}
In Appendix \ref{orw:SurfBun}, we find that the set of allowed
$(\eta_p,\eta_c)$ is exactly given by the integer pairs that satisfy
\begin{align}
 \eta_c \se{8} \eta_p \se{4} 0.
\end{align}
Now \eqn{etaZ} implies that $\nu D_{U_1\ft \Si_g} \in \Z$ and 
two new quantization
conditions
\begin{align}
c  D_{U_1\ft \Si_g} &\in \Z , \ \ \ \ g\geq 5,
\nonumber\\
(-\frac{c}{2}  +  \frac \nu 2) D_{U_1\ft \Si_g} &\in \Z , \ \ \ \ g\geq 5.
\end{align}
To summarize
\frmbox{for a 2+1D fermionic gapped phase with $U^f_1$ symmetry,
its chiral central charge $c$ and filling fraction $\nu$
satisfy the following quantization conditions
\begin{align}
\label{sumF}
c  D_{U_1\ft \Si_g} & = 0 \text{ mod } 1 , \ \ \ \ g\geq 5,
\nonumber\\
 \nu D_{U_1\ft \Si_g} & = 0 \text{ mod } 1, \ \ \ \ g\geq 1,
\nonumber\\
(-c  +  \nu ) D_{U_1\ft \Si_g} & = 0 \text{ mod } 2 , \ \ \ \ g\geq 5,
\end{align}
where $D_{U_1\ft \Si_g}$ is the ground state degeneracy
on closed space $\Si_g$.
}

\subsection{H-type invertible  fermionic $U^f_1$-enriched topological orders}
\label{HinvF}

Let us discuss a simple example.  The invertible fermionic $U^f_1$-enriched
topological order are classified by $K$-matrix (see \eqn{KmatL})
\begin{align}
 K = (1)^{\oplus m} \oplus (-1)^{\oplus n} 
\end{align}
and charge vector $\v q$ with odd-integer components.  Its $c$ and $\nu$ are
given by
\begin{align}
 c=\Tr(K),\ \ \ \
\nu= 
\sum_{i=1}^m q_i^2 -\sum_{i=m+1}^{m+n} q_i^2 . 
\end{align}
The ground state degeneracy is always $D_{U_1\ft \Si_g}=|\Det(K)|=1$.  We see
that $-c+\nu= 0$ mod 8, which satisfies (but does not saturate) \eqn{sumF}.

\section{The topological invariants for L-type topological orders}

\label{cob}

\subsection{Topological partition function for $L$-type topological order}

Consider a bosonic system in gapped liquid phase described by a path integral
on $n$D spacetime $M^n$.  After integrating out the all the dynamical degrees
of freedom, we will obtain a partition function that has the following form
\begin{align}
Z(M^n) = \ee^{- \int_{M^n} \dd^n x\, \veps(x)} Z^\text{top}(M^n)
\end{align}
where $\veps(x)$ is the energy density of the ground state.  The term
$Z^\text{top}(M^n)$ is called the topological partition function and is a
topological invariants of the 2+1D topological order.  To be more precise,
$Z^\text{top}(M^n)$ is a complex function on $\cM_{M^n}$, which is the moduli
space of the spacetime $M^n$.  \ie $\cM_{M^n}$ is a space of metrics
$g_{\mu\nu}$ of a closed manifold.  The $g_{\mu\nu}$'s differ by diffeomorphisms
are equivalent and represent the same point of the moduli space $\cM_{M^n}$.

We like to point out that for certain spacetime topologies, the spacetime $M^n$
must contain world-line of point-like excitations, and/or world-sheet of
string-like excitations,  \etc.  For those certain spacetime topologies
$Z^\text{top}(M^n)=0$ on the corresponding moduli space $\cM_{M^n}$.  For other
spacetime topologies, the  spacetime $M^n$ does not contain any excitations. In
this case $Z^\text{top}(M^n)$ is always non-zero on the corresponding moduli
space $\cM_{M^n}$.  If $Z^\text{top}(M^n)$ was zero at an isolated point (or
isolated lower dimensional subspace of  $\cM_{M^n}$), then a small perturbation
will make $Z^\text{top}(M^n)$ non-zero.  This causes the diverging change in
the effective action $S_{eff}=\ln(Z^\text{top}(M^n))$, which indicates that the
system to be gapless.  Thus\cite{KW1458} \frmbox{on the moduli space of gapped
systems, $Z^\text{top}(M^n)$ is either always zero (when the spacetime must
contain excitations), or always non-zero (when the spacetime is filled by
gapped ground state).} We also like to conjecture that \frmbox{on the moduli
space of gapped systems, $|Z^\text{top}(M^n)|$ is a constant.} Physically,
since the system is gapped, the $Z^\text{top}(M^n)$ should not depend on the
size (\ie the metrics) of the spacetime, we naively expect $Z^\text{top}(M^n)$
to be constant.  However, the such a naively expectation is not totally
correct.  We will see that the phase of $Z^\text{top}(M^n)$ can depend on the
metrics. So we conjecture that $|Z^\text{top}(M^n)|$ is a constant.

The metrics dependence of the phase of  $Z^\text{top}(M^n)$ has a topological
reason: they come from the gravitational Chern--Simons terms $ \om_n(a^{SO})$
that depend on the $SO$ connection $a^{SO}$ of the tangent bundle
\begin{align}
 Z^\text{top}(M^n)=|Z^\text{top}(M^n)|
\ee^{2\pi \ii \int_{M^n} \ka \om_n(a^{SO})} .
\end{align}
The gravitational Chern--Simons terms exist only in spacetime dimensions
$n\se{4} 3$.  

When the tangent bundle of the spacetime is non-trivial, the $SO$ connection
$a^{SO}$ of the tangent bundle is not globally defined, and as the result the
gravitational Chern--Simons action $\int_{M^n} \ka \om_n(a^{SO})$ is not well
defined.  In this paper, we use a cobordism approach trying to define the
difference of two gravitational Chern--Simons actions as
\begin{align}
&\ \ \ \
  \int_{M^n_1} \ka \om_n(a^{SO}) - \int_{M^n_2} \ka \om_n(a^{SO}) 
\nonumber\\
&\equiv  \int_{N^{n+1}} \ka \dd \om_n(a^{SO})
= \int_{N^{n+1}} \ka p(a^{SO})
\end{align}
where $M^1_1$ and $M^n_2$ are the boundary of $N^{n+1}$: $\prt N^{n+1} =
M^n_1\sqcup -M^n_2$, and $ p(a^{SO}) = \dd \om_n(a^{SO})$ is a product of
Pontryagin classes.  Note that $a^{SO}$ in $p(a^{SO})$ is the $SO$ -connection
of the tangent bundle for $N^{n+1}$.  Since $N^{n+1}$ satisfying   $\prt
N^{n+1} = M^n_1\sqcup -M^n_2$ is not unique, the different choices of $N^{n+1}$
may give different $\int_{M^n_1} \ka \om_n(a^{SO}) - \int_{M^n_2} \ka
\om_n(a^{SO})$. However, if the ambiguity is an integer, then $\frac{\ee^{2\pi
\ii \int_{M^n_1} \ka \om_n(a^{SO})}}{\ee^{2\pi \ii \int_{M^n_2} \ka
\om_n(a^{SO})}} $ is well defined.  This leads to a quantization of $\ka$:
\begin{align}
\label{kaQ}
 \int_{N^{n+1}} \ka p(a^{SO}) \in \Z,\ \ \
\prt N^{n+1} =\emptyset .
\end{align}
So according to the cobordism approach, the phase of the partition function
$\ee^{2\pi \ii \int_{M^n} \ka \om_n(a^{SO})}$ can be well defined only when
$\ka$ satisfy a quantization condition \eqn{kaQ}.

\subsection{The quantization of $c$ for 2+1D bosonic
topological orders with non-zero partition function}

Let us apply the above result for 2+1D L-type bosonic topological orders.  Here
we make a crucial assumption that the partition function is non-zero for all
closed \emph{orientable} spacetime manifolds
\begin{align}
\label{gcs3}
Z^\text{top}(M^3)
=|Z^\text{top}(M^3)|
\ee^{\ii \frac{2\pi c}{24} \int_{M^3} \om_3 }
\neq 0.
\end{align}
Not all 2+1D L-type bosonic topological orders satisfy this condition.  But at
least, the 2+1D L-type bosonic invertible topological
orders\cite{KW1458,F1478,K1459} are believed to satisfy this condition.

For 2+1D L-type bosonic topological orders satisfying this condition, their
chiral central charges satisfy
\begin{align}
 \frac{c}{24} \int_{N^4} p_1 \in \Z, \ \ \ \forall \ \ \prt N^{4} =\emptyset .
\end{align}
Since the partition function is non-zero on all closed \emph{orientable}
3-manifold, here we choose $N^4$ to an arbitrary closed \emph{orientable}
4-manifold. This is a key assumption in our cobordism approach.

We know that for all different closed \emph{orientable} 4-manifolds $N^4$, the
set $\{ \int_{N^4} p_1 \}$ is given by $3\Z$.  We find that for those  L-type
bosonic topological orders, their chiral central charges are quantized as
\begin{align}
 \frac c 8 \in \Z.
\end{align}
We conclude that, at least for the 2+1D L-type bosonic invertible topological
orders, their chiral central charges satisfy $c=0$ mod 8.

\subsection{The quantization of $c$ for 2+1D bosonic topological orders with
vanishing partition functions only on non-spin manifolds and for 2+1D fermionic
invertible topological orders}

It was pointed out that a bosonic topological order with emergent fermions must
have vanishing topological partition functions on non-spin manifolds.  Here we
consider a subclass of 2+1D bosonic topological order with emergent fermions,
such that topological partition functions vanishes \emph{only} on non-spin
manifolds.  The topological  partition functions are non-zero on all closed
\emph{orientable spin} manifolds.  We believe that the partition functions for
fermionic invertible topological order satisfy this condition.  Also the
bosonic topological order obtained by gauging the $Z_2^f$ symmetry in fermionic
invertible topological order satisfy this condition.  In this case, the
quantization condition becomes
\begin{align}
 \frac{c}{24} \int_{N^4} p_1 \in \Z, \ \ \ \forall \ \ \prt N^{4} =\emptyset,\ \ \w_2=0 .
\end{align}
\ie we also require $N^4$ to be \emph{orientable} and \emph{spin}. As explained in Appendix \ref{orw:4Man}, the set of integers which may be realized as $\int_N p_1$ for closed spin 4-manifolds $N^4$ is $48\mathbb{Z}$. 
We find that for those simple
L-type bosonic topological orders with emergent fermions, their chiral central
charges are quantized as
\begin{align}
 2 c  \in \Z.
\end{align}
We conclude that, at least for the 2+1D L-type fermionic invertible topological
orders, their chiral central charges satisfy $c = 0$ mod $\frac12$.

\subsection{Framing anomaly}

We have mentioned that the partition function is a function on the moduli space
$\cM_{M^3}$ of different closed spacetime manifolds $M^3$.  In fact, in the
presence of gravitational Chern--Simons term, such a statement is incorrect.
The partition function is actually a function on the moduli space $\cM_{M^3}^F$
of different closed spacetime manifolds $M^3$ \emph{with framing structure}.
In other words, the partition functions for the same manifold but with
different framing structures may have different values.

So, what is a framing structure? Give a $n$-manifold $M^n$ and its tangent
bundle $TM$, its stabilized tangent bundle is given by $TM\oplus \R^\infty$.
The manifold $M^n$ is \emph{framable} if the stabilized tangent bundle is
trivial.  A trivialization of the stabilized tangent bundle (\ie a choice of a
global basis of the stabilized tangent bundle) is a framing structure.  

It turns out that all orientable 3-manifold $M^3$ is framable. This gives us a
way to define the gravitational Chern--Simons term.  If the tangent bundle $TM$
of $M^3$ is non-trivial, the $SO_3$ connection $a^{SO_3}$ for the tangent
bundle $TM$ cannot be globally defined on $M^3$, which makes the gravitational
Chern--Simons term $\int_{M^3} \om_3(a^{SO_3})$ not well defined.  But if we
embed $SO_3$ into $SO_\infty \equiv SO$, then the corresponding $a^{SO}$
connection  can be globally defined on $M^3$ (since the stabilized tangent
bundle is trivial), which makes the gravitational Chern--Simons term $\int_{M^3}
\om_3(a^{SO})$ well defined.  But there are different ways to turn a
$a^{SO_3}$-connection into a globally defined $SO$-connection $a^{SO}$, which
corresponds to different trivializations (\ie different choices of framing
structures).  Such different choices of framing structures can change the
gravitational Chern--Simons term $\int_{M^3} \om_3(a^{SO})$ by an arbitrary
integer.  We see that if $c\neq 0$ mod 24, then the phase of partition function
$ \ee^{\ii \frac{2\pi c}{24} \int_{M^3} \om_3(a^{SO}) } $ will depend on the
framing structures and the  partition function is not a function on the moduli
space $\cM_{M^3}$ of 3-manifolds.  This phenomenon is called framing anomaly.
In this case, the  partition function is a function on the moduli space
$\cM_{M^3}^F$ of 3-manifolds with framing structures.

In general, H-type topological orders have $c\neq 0$ mod 24, and thus
correspond to L-type topological orders with framing anomaly.  The L-type
topological orders without framing anomaly must have $c\se{24} 0$.  In this
paper, we will mainly discuss L-type topological orders with framing anomaly.
Their partition function is a function on the moduli space $\cM_{M^3}^F$ of
3-manifolds with framing structures.

\section{The topological invariants for L-type topological orders with $U_1$
symmetry}

\subsection{Quantization of $c$ and $\nu$ for bosonic 
$U_1$-SET orders with non-zero partition functions}

A L-type bosonic $U_1$-enriched topological orders may contain two Chern--Simons
terms given by \eqn{Leff}.  Here we assume that the partition function is
non-zero for all closed \emph{orientable} spacetime with any $U_1$-bundle on it.
Repeating the arguments  in the last section, but for spacetime with a $U_1$
connection $a^{\RZ}=\frac{A_{eff}}{2\pi}$, we obtain the following quantization
for $c$ and $\nu$, for bosonic $U_1$-enriched topological orders whose
topological partition functions are non-zero on any closed orientable
spacetime:
\begin{align}
\int_{N^4}  
-\frac{c}{24} p_1 
+\frac{\nu}{2} c_1^2
\in \Z, \ \ \ \forall \  \prt N^{4} =\emptyset \text{ and $\forall \ U_1$ bundles}.
\end{align}
Here $N^4$ is an arbitrary \emph{orientable} 4-manifold with an arbitrary
$U_1$-bundle on it.  Let $3\eta_p = \int_{N^4} p_1$ and $\eta_c = \int_{N^4}
c_1^2$, where $\eta_p$ and $\eta_c$ are integers.
In Appendix \ref{orw:4Man} we show that the pairs $(\eta_p,\eta_c)$ which may be realized is given by $(\Z,\Z)$.  This leads to the quantization of $c$ and $\nu$:
\begin{align}
 c = 0 \text{ mod } 8, \ \ \ \ \ \nu = 0 \text{ mod } 2.
\end{align}
The above result, at least, applies to L-type invertible bosonic $U_1$-enriched
topological orders and L-type bosonic $U_1$-SPT
orders .

\subsection{Quantization of $c$ and $\nu$ for fermionic $U_1^f$-enriched
topological orders with non-zero partition functions}

A L-type fermionic $U_1^f$-enriched topological orders also contain two
Chern--Simons terms given by \eqn{Leff}.  For fermionic $U_1$-SET orders whose
topological partition functions are non-zero on any closed smooth
\emph{orientable spin$^\C$} spacetime manifolds, the quantization for $c$ and
$\nu$ is given by
\begin{align}
\int_{N^4}  
-\frac{c}{24} p_1 
+\frac{\nu}{8} (c_1^{U_1})^2
\in \Z, \ \ \ \forall \  \prt N^{4} =\emptyset \text{ and }
\forall \ c_1^{U_1} \se{2,\dd} \w_2.
\end{align}
Here $N^4$ is an arbitrary closed smooth \emph{orientable spin$^\C$}
4-manifold, as implied by the condition $ c_1^{U_1} \se{2,\dd} \w_2$.  
Let $3\eta_p =
\int_{N^4} p_1$ and $\eta_c = \int_{N^4}  (c_1^{U_1})^2$, where $\eta_p$ and
$\eta_c$ are integers.
As explained in Appendix \ref{orw:4Man}, we may find such spin$^\C$ 4-manifolds realizing any integers $\eta_p$ and $\eta_c$  as long as they satisfy
\begin{align}
 \eta_p\se{8} \eta_c . 
\end{align}
This leads to the quantization of $c$ and $\nu$:
\begin{align}
\label{cnuU1f}
 c &= 0 \text{ mod } 1, \ \ \ \ \ \nu = 0 \text{ mod } 1 ,
\nonumber\\
  -c  + \nu &= 0 \text{ mod } 8 .
\end{align}
The above result, at least, applies to L-type invertible fermionic
$U_1^f$-enriched topological orders and L-type fermionic $U_1^f$-SPT orders.  

For L-type fermionic $U_1^f$-SPT orders, the central charge vanishes $c=0$ and
$\nu=0$ mod 8.  This agrees with the result in \Ref{W1447}.  We also note that
the fermionic invertible $U_1^f$-SET states discussed in Sec. \ref{HinvF}
satisfy the above quantization condition, and actually saturate the
quantization condition.

We remark that the combination of invertible fermionic $U_1^f$-enriched
topological orders and fermionic $U_1^f$-SPT orders (\ie the invertible
fermionic gapped liquid phases with $U^f_1$ symmetry) is classified via
spin$^\C$ cobordism in \Ref{FH160406527}, which is given by $\Z^2$ in 2+1D.
This agrees with our result that those state are labeled by two integers $\{
(\eta_p,\eta_c)\ |\ \eta_p\se{8} \eta_c \}$.  However, the values of $c$ and
$\nu$ are not discussed in  \Ref{FH160406527}.  

\section{Summary}

For 2+1D gapped liquid phases with $U_1$ symmetry, the central charge $c$ and
the dimensionless Hall conductance $\nu$ are described by the Chern--Simons
terms in the partition function.  However, for arbitrary smooth spacetime
3-manifolds and for arbitrary $U_1$-bundle over spacetime 3-manifolds, it is
highly non-trivial to define the Chern--Simons terms.  In this paper, we use a
cobordism approach to define the Chern--Simons terms.  We find that for H-type
quantum systems, the 4-manifolds used in the cobordism approach must be a
surface bundle.  For L-type quantum systems, the 4-manifolds must have the same
type as that of the spacetime 3-manifolds, such that the partition functions
are non-zero.  This leads to different quantization conditions for $c$ and
$\nu$.  In particular, for the H-type quantum systems, the quantization of $c$
and $\nu$ depends on the ground state degeneracies on Riemannian
surfaces.  While for the L-type quantum systems, the quantization of $c$ and
$\nu$ depends on the type of spacetime manifolds where the topological partition
function is non-zero.

~

~

ORW is supported by the ERC under the European Union's Horizon 2020 research and innovation programme (grant agreement No.\ 756444), and by a Philip Leverhulme Prize from the Leverhulme Trust.
LT is supported by the Croucher Fellowship for Postdoctoral Research. 
XGW is partially supported by NSF DMS-1664412 and by the Simons Collaboration on
Ultra-Quantum Matter, which is a grant from the Simons Foundation (651440).

\appendix

\section{Characteristic numbers of 4-manifolds}
\label{orw:4Man}

The most fundamental invariant of a closed oriented 4-manifold $M$ is its intersection form
\begin{align*}
Q_{M} : &\,\, H^2(M^4;\Z)\times H^2(M^4;\Z) \to \Z\\
&(a,b) \mapsto \<a\smile b,[M^4]\>=\int_{M} ab.
\end{align*}
Under connected sum the intersection form satisfies 
\begin{align*}
 Q_{M\#N} = Q_{M} \oplus Q_{N}.
\end{align*}
By Poincar\'e duality it is unimodular, and by commutativity of the cup-product it is symmetric and therefore has a signature, denoted $\sigma(M)$. By a theorem of Thom, $M$ is the boundary of a 5-manifold if and only if its signature is 0.

Hirzebruch's signature theorem relates the signature to Pontrjagin classes, and for a 4-manifold gives
\begin{align}\label{B1}
\int_M p_1(TM) = 3 \cdot \sigma(M).
\end{align}

By definition of the Wu class $u_2$ for any $\mathbb{Z}_2$-cohomology class $x$ we have $\int_M x^2 = \int_M x \cdot u_2$. As $M$ is oriented,
so its first Stiefel--Whitney class vanishes, and $u_2 \equiv
\w_2+\w_1^2=\w_2$, so
\begin{align}\label{B2}
\int_M x^2 = \int_M x \cdot \w_2.
\end{align}
 Thus if $M$ is spin, i.e.\ $\w_2=0$, then the form $Q_M$ is even. If $M$ is spin then by Rochlin's theorem its signature is divisible by 16, and so
\begin{align}\label{B1spin}
\int_M p_1(TM) \se{48} 0.
\end{align}

Finally consider spin$^\C$ 4-manifolds $M$, with $c_1 \in H^2(M;\mathbb{Z})$ the associated Chern class. Then $c_1$ reduces modulo 2 to $\w_2$, so is a characteristic element of $Q_M$. In this case we may also form the characteristic number $\int_M c_1^2$. By an elementary property of symmetric forms over $\mathbb{Z}$ (see II.5.2 of \cite{MilnorHusemoller}) we have
\begin{align}\label{B4}
\int_M c_1^2 \se{8} \sigma(M).
\end{align}
We now describe to what extent these characteristic numbers may be realized.

\vspace{1ex}

\noindent\textbf{Oriented 4-manifolds.} The manifold $\mathbb{C}P^2$ has $Q_{\mathbb{C}P^2}$ given by the 1-by-1 matrix $(1)$, so has signature $1$; similarly $\overline{\mathbb{C}P}^2$ has signature $-1$. By taking connected sums we may therefore realize every element of $\mathbb{Z}$ as the signature of an oriented 4-manifold, so may realize every element of $3\mathbb{Z}$ as $\int_M p_1(TM)$.

\vspace{1ex}

\noindent\textbf{Oriented 4-manifolds with $U_1$-bundle.} The manifold ${\mathbb{C}P}^2 \# \overline{\mathbb{C}P}^2$ has signature 0, and cohomology ring $H^*({\mathbb{C}P}^2 \# \overline{\mathbb{C}P}^2; \mathbb{Z}) = \mathbb{Z}[x, y]/(x^2 + y^2, x^3, y^3)$. Taking the $U_1$-bundle with $c_1 = x$ it has $\int_{{\mathbb{C}P}^2 \# \overline{\mathbb{C}P}^2} c_1^2 = 1$, and taking instead the $U_1$-bundle with $c_1 = y$ it has $\int_{{\mathbb{C}P}^2 \# \overline{\mathbb{C}P}^2} c_1^2 = -1$. By forming connected sums of $\mathbb{C}P^2$ and $\overline{\mathbb{C}P}^2$ with trivial $U_1$-bundle, and ${\mathbb{C}P}^2 \# \overline{\mathbb{C}P}^2$ with the $U_1$-bundles just described, we can realize any element of $\mathbb{Z}^2$ as $(\sigma(N), \int_N c_1^2)$ for an oriented 4-manifold $N$ with a $U_1$-bundle over it.
\vspace{1ex}

\noindent\textbf{Spin 4-manifolds} The 4-manifold $K3$ is spin, and has intersection form
\begin{align*}
 Q_{K3} = -E_8^{\oplus 2} \oplus H^{\oplus 3},
\end{align*}
where
\begin{align}
\label{E8}
H = \begin{pmatrix}
 0&1\\ 1&0\\
\end{pmatrix},
\ \ \ \
E_8 =
 \begin{pmatrix}
2&1&0&0&0&0&0&0\\ 
1&2&1&0&0&0&0&0\\ 
0&1&2&1&0&0&0&0\\ 
0&0&1&2&1&0&0&0\\
0&0&0&1&2&1&0&1\\ 
0&0&0&0&1&2&1&0\\ 
0&0&0&0&0&1&2&0\\ 
0&0&0&0&1&0&0&2\\
\end{pmatrix}
\end{align}
so has signature $-16$. By taking connected sums of $K3$ and its orientation-reversed analogue $\overline{K3}$ we may realize every element of $16\mathbb{Z}$ as the signature of a spin 4-manifold, so may realize every element of $48\mathbb{Z}$ as $\int_M p_1(TM)$.

\vspace{1ex}

\noindent\textbf{Spin$^\C$ 4-manifolds.} The manifold $\mathbb{C}P^2$ has cohomology ring $H^*(\mathbb{C}P^2; \mathbb{Z}) = \mathbb{Z}[x]/(x^3)$. Its usual complex structure induces a spin$^\C$-structure having $c_1 = 3x$, and so having
\begin{align*}
\sigma(\mathbb{C}P^2) &= 1\\
\frac{1}{8}\left(\sigma(\mathbb{C}P^2) - \int_{\mathbb{C}P^2} c_1^2 \right) &= -1.
\end{align*}
The manifold $\mathbb{C}P^1 \times \mathbb{C}P^1$ has cohomology ring $H^*(\mathbb{C}P^1 \times \mathbb{C}P^1; \mathbb{Z}) = \mathbb{Z}[x, y]/(x^2, y^2)$. Its usual complex structure induces a spin$^\C$-structure having $c_1 = 2x + 2y$, and so having
\begin{align*}
\sigma(\mathbb{C}P^1 \times \mathbb{C}P^1) &= 0\\
\frac{1}{8}\left(\sigma(\mathbb{C}P^1 \times \mathbb{C}P^1) - \int_{\mathbb{C}P^1 \times \mathbb{C}P^1} c_1^2\right) &= -1.
\end{align*}
Therefore by taking connected sums, and changing orientations, there are spin$^\C$ 4-manifolds $M$ realising any values of $\sigma(M)$ and $\int_M c_1^2$ satisfying \eqref{B4}.

\section{Characteristic numbers of surface bundles over surfaces}
\label{orw:SurfBun}

Let $\pi : E^4 \to B^2$ be a smooth fiber bundle, with $E$ and $B$ closed oriented manifolds and fiber the oriented genus $g$ surface $\Sigma_g$. Suppose we are given a $U_1$-bundle on $E$, or equivalently a class $c_1 \in H^2(E;\mathbb{Z})$. This appendix describes the various characteristic classes and numbers which arise in this situation, and then explains the extent to which these may be realized.

\subsection{Definitions and relations}\label{orw:DefCharClass}

Let $T_\pi E = \mathrm{Ker}(D\pi : TE \to \pi^*TB)$ denote the subbundle of the tangent bundle of $E$ consisting of tangent vectors parallel to the fibers of $\pi$, giving a decomposition $TE = T_\pi E \oplus \pi^*TB$. The tangent bundle of an orientable surface, such as $B$, becomes trivial upon the addition of a trivial line bundle, and hence its Stiefel--Whitey and Pontryagin classes vanish. We therefore have
\begin{align*}
p_1(TE) = p_1(T_\pi E) & =: p_1 \in H^4(E;\mathbb{Z})\\
\w_2(TE) = \w_2(T_\pi E) & =: \w_2 \in H^2(E;\mathbb{Z}_2).
\end{align*}
In addition, the orientations of $TE$ and $TB$ induce an orientation of the 2-dimensional bundle $T_\pi E$, and we have its Euler class
$$e(T_\pi E) =: e \in H^2(E;\mathbb{Z}).$$
As for any oriented 2-dimensional bundle, we have the identities
\begin{align}
p_1 &= e^2 \in H^4(E;\mathbb{Z})\label{G1}\\
r_2(e) &= \w_2 \in H^2(E;\mathbb{Z}_2),\label{G2}
\end{align}
where $r_2 : H^*(- ; \mathbb{Z}) \to H^*(- ; \mathbb{Z}_2)$ denotes reduction modulo 2. 

Using these classes we may form $\int_E p_1$, which by \eqref{G1} agrees with $\int_E e^2$,  known as the first Miller--Morita--Mumford class. By \eqref{B1} we have $\int_E p_1 = 3 \cdot \sigma(E)$ so $\int_E p_1$ is always divisible by 3, but in fact more is true: as a consequence of the Atiyah--Singer index theorem applied to the fiberwise signature operator we have
\begin{align}\label{G4}
\int_E p_1 \se{12} 0.
\end{align}
This may be obtained from p.\ 555 of \Ref{MoritaChar}, using that $B_2 = \tfrac{1}{6}$. If in addition $\w_2=0$, then \eqref{B1spin} gives
\begin{align}\label{G4bis}
\int_E p_1 \se{48} 0.
\end{align}

Now let us suppose that we are further given a $c_1 \in H^2(E;\mathbb{Z})$. Using
this we may form $\int_E c_1^2$ and $\int_E e \cdot c_1$.  As $e$ reduces modulo 2 to $\w_2$ by \eqref{G2}, \eqref{B2} gives
\begin{align}\label{G5}
\int_E c_1^2 \se{2} \int_E e \cdot c_1.
\end{align}

Let us now suppose in addition that $r_2(c_1) = \w_2$, i.e.\ that $c_1$ provides a fiberwise spin$^\C$-structure, and investigate its consequences. As the Euler class $e$ also reduces to $\w_2$ modulo 2, it follows from the Bockstein sequence that $c_1 = e + 2 \bar{c}_1$ for some $\bar{c}_1 \in H^2(E;\mathbb{Z})$. Then we have $\int_E c_1^2 = \int_E e^2 + 4 \int_E (e \bar{c}_1 + \bar{c}_1^2)$ which by the divisibility results above we can write in terms of integers as $\int_E c_1^2 = 12 \cdot\left(\frac{\int_E p_1}{12}\right) + 8 \cdot\left(\frac{\int_E (e \bar{c}_1 + \bar{c}_1^2)}{2}\right)$. This implies
\begin{align}\label{G6}
\int_E c_1^2 \se{8} \frac{\int_E p_1}{3},
\end{align}
recovering \eqref{B4}. Similarly, $\int_E e \cdot c_1 = 12 \cdot\left(\frac{\int_E p_1}{12}\right) + 2 \cdot \int_E e \cdot \bar{c}_1$ so we have
\begin{align}\label{G7}
\int_E e \cdot c_1 \se{2} 0.
\end{align}

\subsection{Realising characteristic numbers}\label{orw:Realising}

The Madsen--Weiss theorem \cite{MW} and its variants \cite{CM, GalatiusSpin} can be used to establish the existence of surface bundles with given geometric structure and characteristic numbers in a highly indirect way, assuming that the genus $g$ is sufficiently large in comparison with the dimension of the base. For the structure considered here, of a surface bundle equipped with a $U_1$-bundle (equivalently a second integral cohomology class) on the total space and 2-dimensional base, this has been carried out in \cite{ERW}. The conclusion is as follows.

Firstly, for any $g \geq 5$ there are oriented $\Sigma_g$-bundles $\pi : E^4 \to B^2$ and classes $c_1 \in H^2(E;\mathbb{Z})$ which realize any values of
\begin{align*}
\int_{\Sigma_g} c_1,\ \int_E p_1,\ \int_E c_1^2,\ \int_E e \cdot c_1
\end{align*}
as long as they satisfy \eqref{G4} and \eqref{G5}. This is obtained by combining Theorem A (iv) and Theorem C of \Ref{ERW} and using that every second homology class may be represented by a map from a surface $B^2$.

Secondly, for any $g \geq 5$ there are oriented $\Sigma_g$-bundles $\pi : E^4
\to B^2$ and classes $c_1 \in H^2(E;\mathbb{Z})$ satisfying $r_2(c_1)=\w_2$ which
realize any values of
\begin{align*}
\int_{\Sigma_g} c_1,\ \int_E p_1,\ \int_E c_1^2,\ \int_E e \cdot c_1
\end{align*}
as long as they satisfy \eqref{G4}, \eqref{G6}, and \eqref{G7}. This is
obtained by taking $c_1 = e + 2 \bar{c}_1$ and translating the conditions
on $\int_{\Sigma_g} \bar{c}_1, \int_E p_1, \int_E \bar{c}_1^2, \int_E e \cdot
\bar{c}_1$ from the previous paragraph to conditions on $\int_{\Sigma_g} c_1, \int_E p_1 \int_E c_1^2, \int_E e \cdot c_1$.

Thirdly, the analogous analysis for spin surface bundles though not equipped with a additional line bundle, has been carried out in \cite{RWSpinPic}. It follows from Example 1.11 of loc.\ cit.\ that for any $g \geq 9$ there are Spin $\Sigma_g$-bundles $\pi : E^4 \to B^2$ realising any value of $\int_E p_1$ satisfying \eqref{G4bis}.

\section{Checking \eqref{gsdCsu2}, \eqref{nuD1}, \eqref{nuD2}} \label{su2cs}

\subsection{Spread polynomials $S_n(x)$}

The spread polynomials $S_n(x)$ are defined by $S_n(\sin^2\th)=\sin^2(n\th)$.
They have the explicit form\cite{WJ05}:
\begin{align*}
S_n(x) = x\sum_{m=0}^{n-1}\frac{n}{n-m}{2n-1-m\choose m }(-4x)^{n-m-1}
\end{align*}
which has roots $\{\sin^2(\frac{m\pi}{n}):m=0,\dots,n-1\}$. Putting $n=k+2$ and replacing $x\ra1/x$, the polynomial
\begin{align*}
&x^{k+2}S_{k+2}(\frac{1}{x}) = \sum_{m=0}^{k+1} \al_m x^m \nn
&\al_m:= \frac{k+2}{k+2-m}{2k+3-m\choose m }(-4)^{k-m+1}
\end{align*}
has roots $\{\csc^2\frac{m\pi}{k+2}:m=1,\dots,k+1\}$.
\subsection{Symmetric polynomials $e_l$ and $p_l$}
Let $\xi_m := \csc^2\frac{m\pi}{k+2}$. Their elementary symmetric polynomials are given by
\begin{align*}
e_{l}:&=\sum_{1\le m_1<\dots<m_l\le k+1} \xi_{m_1}\dots\xi_{m_l} = \frac{(-1)^l\al_{k-l+1}}{\al_{k+1}}\nn
&=\frac{4^l}{(k+2)(l+1)}{k+2+l \choose 2l+1}
\end{align*}
for $l\le k+1$. For $l>k+1$, $e_{l}:=0$. Define
\begin{align*}
E_l:&=\frac{(k+2)^l}{2^l}e_l = 2^l(k+2)^{l-2}\frac{k+2}{l+1}{k+2+l\choose 2l+1}\nn
&=2^l(k+2)^{l-2}\big[{k+l+3\choose 2l+2}+{k+l+2\choose 2l+2}\big].
\end{align*}
Note that $E_1 = {k+3 \choose 3}\in\Z$, which is odd iff $k=0 \text{ mod }4$, so $kE_l\in 4\Z$. For $l\ge2$, $E_l\in 4\Z$. Moreover $\frac{3E_2}{k+2}\in 2\Z$ and $\frac{E_{l}}{k+2}\in 2\Z$ for $l\ge 3$.

Define the $l$-th power sums $p_l := \sum_{m=1}^{k+1} \xi_m^l$. They are related to $e_l$'s via Newton--Girard Formulas\cite{SR00}:
\begin{align}
p_l = (-1)^{l+1}l e_{l} + \sum_{j=1}^{l-1}(-1)^{j+1}e_j p_{l-j}.\label{ngf}
\end{align}

\subsection{Checking \eqref{gsdCsu2}}
Define $n_l = \frac{3k}{k+2} \frac{(k+2)^l}{2^l}p_l$. The expression \eqn{gsdCsu2} to be checked is 
\begin{align*}
n_{g-1}\in2\Z.
\end{align*}
We will show this by induction for $g\ge3$. From \eqref{ngf} we have
\begin{align}
n_l = (-1)^{l+1} lk \big(\frac{3E_l}{k+2}\big) + \sum_{j=1}^{l-1}(-1)^{j+1}E_j n_{l-j}.\label{nl}
\end{align}
Note that $n_1=\frac{(k+3)(k+1)k}{2}\in\Z$, which is odd iff $k=2\text{ mod }4$. For $l=2$, we have
\begin{align*}
n_2 =-2k \big(\frac{3E_2}{k+2}\big) + E_1 n_1.
\end{align*}
where the first term is even since the factor in bracket is even as discussed before. The second term is even since $E_1$ and $n_1$ cannot be odd at the same time. For $l\ge3$, it can be seen that $n_l\in 2\Z$ inductively since both terms of \eqref{nl} are even integers.

\subsection{Checking \eqref{nuD1}}
Define $m_l = k \frac{(k+2)^l}{2^l}p_l$. The expression \eqn{nuD1} to be checked is 
\begin{align*}
m_{g-1}\in2\Z.
\end{align*}
We will show this by induction for $g\ge1$. From \eqref{ngf} we have
\begin{align}
m_l = (-1)^{l+1} lkE_l + \sum_{j=1}^{l-1}(-1)^{j+1}E_j m_{l-j}.\label{ml}
\end{align}
Note that $m_0=k(k+1)\in2\Z$. For $l\ge1$, the first term in \eqref{ml} is even since $kE_l$ is even. Hence $m_l\in2\Z$ for all $l\ge0$ by induction.

\subsection{Checking \eqref{nuD2}}
The expression \eqn{nuD2} to be checked is 
\begin{align*}
\frac{m_{g-1}}{2}\in2\Z.
\end{align*}
for $g\ge6$. We will show a stronger result that this holds for all $g\ge2$ by induction. \eqref{ml} can be rewritten as
\begin{align}
\frac{m_l}{2} &= (-1)^{l+1} l\big(\frac{kE_l}{2}\big) + \sum_{j=1}^{l-1}(-1)^{j+1}E_j \big(\frac{m_{l-j}}{2}\big)\nn
\frac{m_l}{2}&\se{2}\sum_{j=1}^{l-1} E_j \big(\frac{m_{l-j}}{2}\big)\se{2}
\begin{cases}
E_1\big(\frac{m_{l-1}}{2}\big) &\text{for }l\ge2\\
0 &\text{for }l=1.
\end{cases} \label{ml2}
\end{align}
where we used the fact that $\frac{kE_l}{2}\in 2\Z$ for $l\ge1$ and $E_j\in4\Z$ for $j\ge2$. Note that $\frac{m_0}{2}=\frac{k(k+1)}{2}$ is odd iff $k=1,2$ mod 4. Also recall that $E_1\in\Z$. It is straightforward to read off that $\frac{m_1}{2}\in2\Z$ and hence $\frac{m_l}{2}\in2\Z$ for all $l\ge1$ which follows by induction.

\section{Group extension and trivialization}

\label{cenext}

Consider an extension of a group $H$
\begin{align}
 A \to G \to H
\end{align}
where $A$ is an Abelian group with group multiplication given by
$x+y \in A$ for $x,y\in A$. 
Such a group extension is denoted by $G=A\ft H$.  
It is
convenient to label the elements in $G$ as $(h,x)$, where $h\in H$ and $x
\in A$.  The group multiplication of $G$ is given by
\begin{align}
 (h_1,x_1)(h_2,x_2)=(h_1h_2,x_1+\al(h_1)\circ x_2+e_2(h_1,h_2)) .
\end{align}
where $e_2$ is a function 
\begin{align}
 e_2: H\times H \to A,
\end{align}
and $\al$ is a function
\begin{align}
 \al: H \to \text{Aut}(A).
\end{align}
We see that group extension is defined via $e_2$ and $\al$.
The associativity 
\begin{align}
&
[(h_1,x_1)(h_2,x_2)](h_3,x_3)
=(h_1,x_1)[(h_2,x_2)](h_3,x_3)]
\end{align}
requires that
\begin{align}
&
x_1+\al(h_1)\circ x_2+e_2(h_1,h_2)+\al(h_1h_2)\circ x_3+e_2(h_1h_2,h_3)
\nonumber\\ 
&=\al(h_1)\circ x_2+\al(h_1)\al(h_2)\circ x_3+\al(h_1)\circ e_2(h_2,h_3)]
\nonumber\\ & \ \ \ \
+x_1+e_2(h_1,h_2h_3)
\end{align}
or
\begin{align}
 \al(h_1)\al(h_2) = \al(h_1h_2)
\end{align}
and
\begin{align}
& e_2(h_1,h_2)-e_2(h_1,h_2h_3)+e_2(h_1h_2,h_3)
\nonumber\\ & 
-\al(h_1)\circ e_2(h_2,h_3)=0.
\end{align}
Such a $e_2$ is a group 2-cocycle $e_2\in H^2(\cB H;A_\al) $, where $H$ has a
non-trivial action on the coefficient $A$ as described by $\al$.  Also, $\al$
is a group homomorphism $\al: H \to \text{Aut}(A)$.  We see that the $A$
extension from $H$ to $G$ is described by a group 2-cocycle $e_2$ and a
homomorphism $\al$.  Thus we can more precisely denote the group extension by
$G=A\ft_{e_2,\al} H$.

Note that the homomorphism $\alpha: H \to \text{Aut}(A)$ is in fact the
action by conjugation in $G$,
\begin{align}
  (h,0)(\one, x)=(h,\alpha(h)\circ x)&=(\one,\alpha(h)\circ x)(h,0),\nonumber\\
  \Rightarrow (h,0)(\one,x)(h,0)^{-1}&=(\one,\alpha(h)\circ x).
\end{align}
Thus, $\alpha$ is trivial if and only if $A$ lies in the center of $G$. This
case is called a central extension, where the action $\alpha$ will be omitted.

Our way to label group elements in $G$:
\begin{align}
 g=(h,x) \in G
\end{align}
defines two projections of $G$:
\begin{align}
 \pi: G\to H, \ \ \ \pi(g)=h,
\nonumber\\
 \si: G\to A, \ \ \ \si(g)=x.
\end{align}
$\pi$ is a group homomorphism while $\si$ is a generic function.
Using the two projections, $g_1g_2=g_3$ can be written as
\begin{align}
&\ \ \ \
 (\pi(g_1),\si(g_1)) (\pi(g_2),\si(g_2))
\nonumber \\
&=[\pi(g_1)\pi(g_2),\si(g_1)+\al(\pi(g_1))\circ \si(g_2)+e_2(\pi(g_1),\pi(g_2)) ]
\nonumber \\
&= (\pi(g_3),\si(g_3))= (\pi(g_1g_2),\si(g_1g_2))
\end{align}
We see that the group cocycle $e_2(h_1,h_2)$ in $H^2(\cB H;A) $
can be pullback to give a group cocycle $e_2(\pi(g_1),\pi(g_2))$ in $H^2(\cB G;A) $, and such a pullback is a coboundary
\begin{align}
\label{e2sig}
e_2(\pi(g_1),\pi(g_2)) = -\si(g_1) +\si(g_1g_2) -\al(\pi(g_1))\circ \si(g_2),
\end{align}
\ie an element in $B^2(\cB G; A_\al)$,
where $G$ has a non-trivial action on the coefficient $A$ as described by $\al$. 

The above result can be put in another form. Consider the homomorphism
\begin{align}
 \vphi: \cB G \to \cB H
\end{align}
where $G=A\ft_{e_2}H$, and $e_2$ is a $A$-valued 2-cocycle on
$\cB H$.  The homomorphism $\vphi$ sends an link of  $\cB G$ labeled by
$a^{G}_{ij}\in G$ to an link of  $\cB H$ labeled by
$a^{H}_{ij}=\pi(a^{G}_{ij})\in H$.  The pullback of $e_2$ by $\vphi$, $\vphi^*
e_2$, is always a coboundary on $\cB G$

The above discussion also works for continuous group, if we only consider a
neighborhood near the group identity $\one$.  In this case, $e_2(h_1,h_2)$ and
$\al(h)$ are continuous functions on such a neighborhood.  But globally,
$e_2(h_1,h_2)$ and $\al(h)$ may not be continuous functions.

\bibliography{../../bib/wencross,../../bib/all,../../bib/publst,./ORWbib,./local}
\end{document}